# A bibliometric tool to assess the regional dimension of university-industry research collaborations[1]

Giovanni Abramo[a,b,*], Ciriaco Andrea D'Angelo[b] and Marco Solazzi[c]

[a] Institute for System Analysis and Computer Science (IASI-CNR)
National Research Council of Italy

[b] Laboratory for Studies of Research and Technology Transfer
School of Engineering, Dept of Management
University of Rome "Tor Vergata"

[c] NEXT Ingegneria dei Sistemi S.p.A.

**Abstract**

The present work proposes a bibliometric methodology for measuring the grade of correspondence between regional industry's demand for research collaboration and supply from public laboratories. The methodology also permits measurement of the intensity and direction of the regional flows of knowledge in public-private collaborations. The aim is to provide a diagnostic instrument for regional and national policy makers, which could add to existing ones to plan interventions for re-balancing sectorial public supply of knowledge with industrial absorptive capacity, and maximizing appropriability of knowledge spillovers. The methodology is applied to university-industry collaborations in the hard sciences in all Italian administrative regions.

**\* Corresponding author:** Dipartimento di Ingegneria dell'Impresa, Università degli Studi di Roma 'Tor Vergata', Via del Politecnico 1, 00133 Rome - ITALY, tel/fax +39 06 72597362, giovanni.abramo@uniroma2.it

## 1. Introduction

Modern theories of endogenous growth place increasing stress on the role of knowledge in processes of economic development. In fact, competitiveness and growth are increasingly seen as linked with knowledge (Metcalfe, 2002). Thus investments in knowledge, specifically in research and development and education expenditures, show a growth trend in most OECD nations (OECD, 2007). The "triple helix" model of development (Etzkowitz and Leydesdorff, 1998) sees tight interaction between the public research sphere, industrial system and government institutions as the best route to favor innovation and development. Under this model, public-private relationships play a primary role in improving national wellbeing. Ever increasing relevance is attached to certain concepts and issues, such as links between public and private research, "knowledge spillovers", geographic proximity, knowledge diffusion and "appropriability". Jaffe (1989), Acs et al. (2002), Feldman (1994), and Anselin et al. (1997), among others, have furnished empirical studies with strong evidence of the positive relationship between research and development and innovative activity, both in the industrial and public spheres.

Among various modes in which the sought-after technology transfers develop, university-industry collaborations result as being particularly effective. In addition, it has also been demonstrated that geographical proximity (see next section) plays an important role in favoring knowledge spillovers from the public research to the productive system. This is the reason for the current attention of policy makers, both at the national and supra-national level, to "Regional Innovation Systems". In particular, in the current debate on regional EU innovation policies, sub-national regions are increasingly playing a role (Barca, 2009). "It is from intensifying the public–private regional knowledge transfer within a framework of Regional Innovation Systems (RIS) that the EU and national governments are expecting to strengthen the innovating ability of individual private, but also public, enterprises in a medium to long-term perspective" (Gerstlberger, 2004).

The present study is inserted in an ongoing theme of investigations concerning the regional dimension of public-private research collaboration.

Recognizing the importance of geographical proximity between the producers and users of new knowledge, meaning between public research institutions and private enterprises, it seems reasonable for regional policy to include objectives for stimulating interaction between these two sides, so as to maximize potential local benefits. Preparatory to the formulation of any policy or stimulus initiatives for public-private research collaboration, there must be accurate analysis of private sector demand and public sector supply of knowledge occurring in a region. However, direct investigations of research collaboration at regional levels between public laboratories and private enterprises results as impractical, due to the difficulty of acquiring pertinent data. Even when data seems available, the resulting analysis is frustrated when it attempts to reconcile the different classifications of scientific competence on supply (such as listed in Thomson Reuters' "Web of Science" scientific categories) with the available sectorial classifications describing industrial demand (such as seen in SIC or Ateco). Given the difficulties of acquiring and "fitting" data, empirical research has generally been limited to few technological-scientific sectors, few research institutions, or is conducted on the basis of sampling surveys. Follow up evaluation of the impact of any policies adopted also proves challenging, due to the difficulty in ascertaining variations in intensities of collaboration between the actors involved: for this again, accurate mapping is completely lacking.

Through application of a bibliometric approach, the present work proposes to contribute to overcome these difficulties and provide a supporting tool for regional policy makers. Together with other investigation tools, this one will assist them to identify and quantify supply and demand for new knowledge, at the sectorial level, in the public-private research "market". Regional policy makers can thus better ascertain the match of knowledge supply to demand, formulate policies for improvements in the structural fit, and evaluate the effectiveness of the actions undertaken to increase public-private collaboration, with reference to the individual sectors that are strategic and significant for local regional planning. The bibliometric instrument proposed has been applied to the case of the 20 Italian administrative regions. For each region, measure is made of the extent to which public supply of knowledge corresponds to private demand (research collaborations observed), in 183 technical-scientific sectors. Measure is also made, for each technical-scientific sector and at the aggregate level, of the relative ability of each region to maximize the economic benefits obtained from the knowledge produced by local universities. The objective of this work is to present a complement methodology and to show its potential through the application to the Italian case; it is neither to recommend regional policies on the basis of the results of its application, nor to interpret the results.

The work is presented as follows: after a brief review of literature on the geographic dimension of diffusion of new knowledge, Section 3 of this study then describes the methodological approach and the dataset used. Section 4 then presents a first aggregate analysis of university-industry collaborations in Italy. The subsequent sections provide deeper analysis, taking the sectorial element into consideration: Section 5 deals with analysis at the single sector level while section 6 continues the exploration through an aggregation at regional level of the sectorial results. The closing section presents the study conclusions and indicates possible directions for future research.

## 2. The geographic dimension of knowledge spillovers

The reinforcement of cooperation between universities and private enterprise has become one of the most frequently used strategies for favoring innovation processes (OECD, 2007). Networks of universities and private enterprises are able to generate impacts on regional innovative capacity not only within the immediate industry concerned, but also in "proximate" industries (Rondé and Hussler, 2005), thus providing significant influence on regional economic productivity (Mueller, 2006). However, public-private knowledge spillovers are limited by distance, especially when they concern tacit knowledge (Polanyi 1985), because such transfers require direct interaction between individuals and therefore a limit on the distance between the people involved (Anselin et al., 1997). It seems that successful innovative processes have a strong characteristic of interaction between actors in local proximity. It also seems that, in this dimension, there are more intense collaboration processes involving the transfer of technical-scientific knowledge from the sphere of public research to that of industrial research (Jaffe, 1989; Autant-Bernard, 2001; Moreno et al., 2005; Parente and Petrone, 2006).

Empirical research demonstrates that the flow of knowledge from public to private sector diminishes with geographic distance (Arundel and Geuna, 2004). In general, the number of collaborations between pairs of partners declines exponentially with increasing distance of separation (Katz, 1994). Greunz (2003) and LeSage et al. (2007) confirm that knowledge spillovers generally develop within the confines of a given region, or at most within single nations. Knowledge flows originating from contiguous regions improve regional growth performance, but their effect declines rapidly with distance (Rodrìguez-Pose and Crescenzi, 2008). Since knowledge spillovers favor technological change and economic growth, research into the modes of knowledge diffusion in relation to geographic factors could assist in understanding the observed variations in level of growth and development that occur between different regions (Audretsch and Keilbach, 2004; Varga and Schalk, 2004).

In the study of ties between proximity and innovation, Boschma (2005) evaluates geographic proximity conjointly with other dimensions of proximity (cognitive, organizational, social and institutional). Geographic proximity can compensate for the absence of other forms of proximity, thus permitting collaborations between organizations of diverse background, mission, organizational structure, etc. (in this case between universities and private companies). Equally, other types of proximity can favor collaboration between actors located at great distances from one another. However, while geographic proximity should increase the possibility of personal interactions that are useful for transfer of tacit knowledge, it is not a sufficient condition for development of collaborations (Breschi and Lissoni, 2001; Moreno et al., 2005).

It is certain that the sectorial element also plays an important role. Research externalities are not uniform, and actually present significant differences among scientific sectors in terms of their impact on public-private research agglomeration phenomena. It has also been observed that the distributions of knowledge spillovers do not have a uniform character and can differ notably, not only for nations and territories, but also in connection with industrial sectors (Anselin et al., 2000). Maggioni and Uberti (2005) arrive at the same conclusions. They have demonstrated that, although geographic distance remains a relevant factor in determining the structure of inter-regional knowledge flows, such flows are favored when regions share similar scientific, technological and sectorial characteristics.

For the successful absorption of new knowledge generated in the public sphere, the processes of diffusion require that the industrial sector have adequate capacities to understand, internalize and develop the knowledge (Cohen and Levinthal, 1990). As a consequence, different regions naturally present different capacities for exploiting research from the public sector to generate innovation.

Audretsch and Lehmann (2005) have shown that the choice of an enterprise to locate near a university bears a positive correlation to the "knowledge capacity" of the surrounding region and to the actual knowledge output of the university. Universities are seen to be preferred partners for high-tech innovation and venturing, precisely because of their influential role in creating and diffusing new knowledge (Van Looy et al., 2003).

Various techniques and measuring tools have been applied in the study of knowledge spillovers. Jaffe (1989) uses a "Geographic Coincidence Index", which provides a measure of "geographic coincidence of university and industrial research activities within the region", and shows that knowledge spillovers are concentrated in specific sectors. Meanwhile, Audretsch and Feldman (1996), studying the spatial concentration of production, have shown that the tendency for clustering in innovative activity is due more to the role of knowledge spillovers than to geographic concentration of production. Direct studies of knowledge spillovers have also been conducted by studying patent registrations. Using information on the geographic location of inventors, and studying the patterns of how one patent cites others, and vice versa, it has been shown that knowledge spillovers are especially evident at the local level (Jaffe et al. 1993). The measure of the geographic aspect of knowledge externalities has also been attempted through tools for spatial econometrics, different from traditional econometrics in that it includes the consideration of spatial effects (Anselin 1988; Fingleton and López-Baso 2006). The current authors use a bibliometric approach based on scientific publications produced through collaborations between university and industry.

**3. Methodological approach**

This study stems from the recognition of the importance of public-private knowledge spillovers for innovation intensity, and of the relevance of i) geographic proximity; ii) degree of concentration of public research institutions and private enterprises and iii) degree of correspondence between their relative research activities. The aim of the study is then to provide the regional policy maker with an additional instrument for the analysis of demand and supply of knowledge in the region and for evaluation of the impact of policies intended to favor local knowledge spillovers. The instrument was applied to the case of the 20 administrative regions that subdivide the Italian national territory. Knowledge spillovers can occur in different ways. The dimension investigated here is that concerning research collaborations between universities and private enterprises located on Italian territory in the period from 2001 to 2003, for which data were available[2]. To identify such collaborations the study resorts to the proxy of articles with university-industry co-authorship as indexed in the CD-ROM version of the Thomson Reuters Science Citation Index (SCI$^{TM}$). The analysis of publication co-authorship for measuring research collaborations between organizations, presents few limits and numerous advantages. Among the formers, the fact that scientific collaborations do not always lead to publication and that real collaboration does not always exist among scientists (Melin and Persson 1996, Katz and Martin 1997, Laudel 2001); furthermore, bibliometric analysis does not delve into the characteristics of the research at stake and the economic returns of its results. Among the latters, the quantifiable and invariant character of such analysis, and the non-invasive and relatively economical cost of its application. In terms of scope of the field of observation, bibliometric analyses present clear advantages over other empirical methods based on partial surveys. In terms of number of observations bibliometrics overwhelms patent analysis: in Italy the number of patents filed by all universities is notoriously low (Abramo and Pugini, 2005). Using the Espacenet search engine we have identified 284 patents filed by Italian universities between 2001 and 2003[3]. Of these, only 20 were co-

---

[2] It should be noted that in addition to universities, research institutes also contribute to the production of new knowledge, but are not fully considered in this work. The current work is primarily intended to describe a measurement system and provide an example of its application to the Italian case: the results should be interpreted in this sense.

[3] Legislation in 2001 introduced the so called "academic privilege", presumably resulting in additional patents filed by university researchers, but relevant data are not readily available, making the

owned with private companies. Because co-ownership does not necessarily indicate the actual authorship of a patent, the number of university patents stemming from true collaboration with industry may actually be less than 20. A more direct approach would entail a survey of research collaboration agreements between universities and industry, however it may reveal a formidable task, because of lack of centralized databases on the subject. The proposed bibliometric tool, anyway, is meant to complement rather than substitute other diagnostic approaches.

The field of observation for this work consists of all 78 Italian universities and all private companies located in national territory in the period considered. The SCI$^{TM}$ was first used to extract the publications (articles and reviews) authored by organizations based in Italy. From this group, the next step was to select those articles co-authored in collaboration between universities and companies. To do this, the authors had to identify and reconcile the different ways in which the same organization was reported in the SCI$^{TM}$ "address" field for the articles.

Finally, through a "disambiguation" algorithm, each publication was accurately attributed to its respective university authors. Such operation is quite complex. Current bibliometric databases (such as Elsevier's SCOPUS and Thomson Reuters' Web of Science) make it formidable to identify the real author of a publication, because the "authors' list" and the "address list" are not linked together. Moreover only authors' last name and first name initials are reported. As a consequence, any time the address list contains two or more affiliations, one does not know immediately to which one each author belongs. Such technical limitations were addressed and overcome by D'Angelo et al. (2011) who developed a disambiguation procedure to attribute each publication to its academic authors with an error below 3%. For our purposes the attribution of each publication to its academic authors is relevant for classifying the collaboration into a specific discipline. In particular in Italy, the academic regulatory system provides that every university researcher must adhere to a specific declared scientific disciplinary sector (SDS), which in turn is a component of a specified university disciplinary area (UDA). Through this structure it is possible to link each publication to the SDS of its university author.

The field of observation for the current analysis is restricted to the hard sciences, represented by 8 technical-scientific UDA (Mathematics and computer sciences, Physics, Chemistry, Earth sciences, Biology, Medicine, Agricultural and veterinary sciences, Industrial and information engineering)[4], comprising a total of 183 SDS. The dataset thus results as composed of the subset of SCI$^{TM}$-indexed publications for 2001 to 2003, co-authored by at least one university scientist falling in one of these 183 SDS, and at least one private enterprise located on Italian territory. This includes a total of 1,534 such publications produced by a total of 58 universities and 483 private enterprises.

To investigate research collaboration, the resulting dataset was subjected to two distinct levels of analysis[5]. In the first, the collaborations were analyzed at the level of organization (university-enterprise collaboration). By university-enterprise collaboration we mean a research collaboration between a university and a private company, both located on Italian territory, that has resulted in exactly one co-authored publication in

---

identification of joint patents very difficult.

[4] Civil Engineering was not considered because the relevant publications are poorly represented in the SCI$^{TM}$.

[5] For further information see Abramo et al., 2011.

the dataset under consideration. A generic publication realized by *m* universities and *n* private enterprises thus corresponds to *m\*n* university-enterprise collaborations. It results that a single university-enterprise partnership, if it has realized *z* publications in the dataset under consideration, provides *z* university-enterprise collaborations. In the second level of analysis, the collaborations were examined at the level of scientific sector (SDS-enterprise collaboration). The sector of collaboration was identified as being that of the SDS to which the university authors adhere. In fact, it is reasonable to assume that a university researcher carries out her/his work within her/his actual area of specialization. Other approaches would also suffer from still greater limitations: given the vast field of observation, resorting to a survey approach would not be opportune, while identifying the sector as that of the private enterprise would involve lengthy national data searches and would still result in a similar problem, that of identifying the SDS of diversified enterprises. Thus, starting from a co-authored publication, and accepting the above assumption, the number of SDS-enterprise collaborations takes count of both the number of private enterprises involved and also the number of SDS to which the university authors pertain.

Using the above definitions, the 1,534 publications realized by universities and private enterprises in Italy in the triennium under consideration produced a total of 1,983 university-enterprise collaborations and 2,363 SDS-enterprise collaborations[6].

## 4. Characterization of regional demand and supply for university-industry research collaboration in Italy

As a first task towards characterizing the market for public-private research collaborations at the regional level, the 1,983 university-enterprise collaborations were first sorted by region, in function of the location of the authors' employing organizations. It was then possible to characterize the supply of knowledge by universities in each region and study it on three levels of analysis, in relation to its meet with demand: i) intra-regional level: collaborations between universities and enterprises located in the same region; ii) inter-regional level: collaborations between the universities of one region and private enterprises located outside of that region; iii) overall national level (corresponding to the sum of the contributions at the two preceding levels): collaborations between universities of one region and private enterprises located throughout all of Italy. The regional picture was completed by considering, in addition to the supply, the demand on the part of the private enterprises of each region. In this case the analysis was again conducted in reference to 3 cases: intra-regional, inter-regional and national.

It was thus possible to characterize each region in terms of: i) supply of collaboration by its local universities; ii) demand for collaboration by local private enterprises; iii) net difference between demand and supply; iv) regional market share of the local universities (ratio of local collaborations to national demand by private enterprises of the region). Table 1 illustrates the measures of the above indicators, for each region.

[Table 1]

---

[6] A number of publications evidently are co-authored by more than one university/company, and by researchers from different SDSs.

It results that the region of Val D'Aosta does not have universities in the period under consideration, nor any enterprise that has undertaken collaborations, and for this reason the region is not included in subsequent analyses. The university of Molise did not register any collaborations. Those in Calabria collaborated only with extra-regional enterprises. Meanwhile, the region of Basilicata does not present any demand for collaboration in the period considered. The regions in which universities registered the highest number of collaborations are, in order: Lombardy (403, or 20% of the total), Emilia Romagna (298; 15% of total) and Tuscany (203; 11% of total). These high numbers could essentially be due to two factors: i) a greater propensity of the universities in these regions to collaborate with the private sector; ii) a high private intra-regional demand (which would facilitate collaborations through the recognized effect of geographic proximity). From the demand side, the regions where private enterprises collaborated most with universities are Lombardy (769; or 39% of total), Lazio (289;15% of total) and Tuscany (215;11% of total). The calculation of net difference between supply and demand for collaborations shows that Emilia Romagna (102 collaborations), Campania (82) and Umbria (55) are the regions with the greatest net excess in export over import of new knowledge through research collaborations. Lombardy (-366), Lazio (-129) and Piedmont (-13) are instead the regions with the highest net level of knowledge import. These latter three regions are those well known as having the highest level of industrial concentration in Italy, in terms of private R&D expenditures.

A private enterprise may need to resort to research collaborations with universities located in other regions due to various factors, for example due to lack of adequate scientific competencies in the local universities, or their lack of ability or willingness to collaborate. The same may occur for universities, for example Azagra-Caro (2007) shows how, in a region with low absorptive capability[7], certain types of researchers that have a greater tendency to collaborate with industry will do so largely with enterprises of greater size, situated outside their own region. In addition, it must also be recognized that a private enterprise could find itself geographically closer to a university situated in another region, rather than a university in its own region: geographical proximity, which definitely plays a fundamental role in collaboration, would then favor inter-regional collaborations. In our empirical analysis this occurred for a very limited number of firms.

The last indicator present in Table 1 is "regional market share", defined as the ratio between intra-regional supply of collaboration and national demand for collaboration by enterprises of a region. This permits a response, for each region, to the question: "What percentage of the demand for collaboration by local industry is captured by local universities?". The maximum value, equal to 100%, would correspond to the situation (desired by regional policy makers) in which the entire industrial demand of a region is satisfied by local universities, capable of responding in full to the needs for knowledge in local private enterprises. Of the total of the 1,983 research collaborations, 690 (35%) are intra-regional, meaning between universities and enterprises in the same region. The regions that succeed in meeting the highest quota of total demand for collaborations from their internal supply are Marche, Puglia and Umbria, all with 75%. Calabria and Molise do not succeed in realizing any demands for intra-regional collaboration.

---

[7] From Azagra-Caro (2007): "We follow Cohen and Levinthal's (1990) definition of absorptive capacity: ''a limit to the rate or quantity of scientific or technological information that a firm can absorb''. To justify the extension of the concept of absorptive capacity from firms to regions see Niosi and Bellon (2002)".

The mismatch between demand and supply in any region offers macroscopic information that can be analyzed at a sectorial level. In general, such analysis could permit the regional policy maker to indicate possible corrective interventions in favor of balancing local demand with local supply. Such interventions would naturally need to be harmonized and coherent from a central national point of view, and also need to consider the interests of the various stakeholders, including universities, industry and government. The next section provides a more in-depth analysis at the sectorial level.

## 5. Sectorial analysis of degree of correspondence and regional flows

Because of the great variety of possible scientific specializations on offer (from university research) and technological specializations in demand (by private enterprise) in any region, it is opportune to analyze public-private research collaborations, at the level of individual scientific-technological sectors. As described above, analysis of the market at a sectorial level permits policy makers to formulate policies for technological and industrial development in recognition of the actual supply and demand of the region. Time-series analyses of the evolution of collaborations can also permit evaluation of the effectiveness of the policy initiatives adopted. An examination at SDS level makes it possible to measure the degree of correspondence between public and private research activity, and also to identify the extent to which the Italian administrative regions are successful ("virtuous") in attempting to maximize local benefits from knowledge spillovers obtained through research collaboration. Aggregating the regional performance in all sectors, weighted for the relative importance of each sector, it is then possible to arrive at comparative measures of the overall regional performance.

In this section of the study, the reference population consists of all SDS-enterprise collaborations, rather than the university-enterprise collaborations used in the previous section. Research collaborations occur in 141 of the 183 SDS under observation. For each SDS, it is possible to characterize the region in terms of the number of SDS-enterprise collaborations involving local universities or enterprises, both at the intra-regional level (internal market) and extra-regional level (external market). Every region can be characterized both in terms of its supply of knowledge (from universities) and the demand (by private enterprise), in each SDS. Analogous to the previous section, there are three levels of analysis according to the location of the partners: intra-regional, extra-regional and national (total).

### 5.1 Degree of correspondence

The analysis of collaborations permits identification of a very important item of information for the policy maker: the degree of correspondence between demand and supply in each SDS. At the analytical level, this degree of correspondence can be described in at least two ways. The first is as the difference between the number of local university scientists appertaining to a particular SDS and the total number of collaborations activated by private enterprises of the region, for the same SDS (the "surplus"). The second is as a ratio of the same figures: the number of collaborations activated in a specific SDS by enterprises in a region, divided by the number of university scientists adhering to the same SDS.

For the purposes of the analysis, we assume that the average university scientist is able to deal with any theme in her/his pertinent SDS, but satisfy only one demand for collaboration in the period under observation[8]. The figures for numbers of scientists in each SDS, averaged over the three years under consideration, were obtained from the data base of university research personnel maintained by Cineca, a consortium of Italian universities, on behalf of the Ministry of Universities and Research. A value above 1 for this measure of correspondence would indicate a shortage, or insufficient supply, while a value less than 1 would reveal a surplus. The further this indicator differs from a value of one, the less will be the correspondence between enterprise demand for collaboration and public supply[9]. A high value for this ratio suggests a high level of industrial demand with respect to the public supply of knowledge, and therefore, other conditions being equal, it indicates a greater probability that internal demand will be satisfied by extra-regional supply, and not only by supply from local universities. On the other hand, a low value for this ratio indicates an excess of supply with respect to demand, meaning that it is unlikely that economic benefits from the new knowledge will occur only in the source region, and instead that they will probably also fall in other regions. Further, it is reasonable to expect that universities in regions with a high value of national demand per university scientist will obtain (or have a higher probability of obtaining) a lower regional market share of collaborations than universities in regions with a low value of demand per scientist.

As an example, Table 2 presents figures for the analysis of correspondence between supply and demand for all regions of Italy in the "Electronics" SDS.

[Table 2]

In the period under observation this particular SDS was the one, among all SDS, that registered the greatest number of articles with co-authorship between Italian universities and enterprises (114 publications), representing a total of 134 SDS-enterprise collaborations. It can be observed that in 7 regions out of 19 there is no demand for research collaboration in the Electronics sector (column 3). However, all but two of these seven regions (Basilicata and Molise) register a supply of university researchers for this SDS (column 2). Thus, in the remaining 5 regions of this group, knowledge spillovers produced in Electronics can not give rise to benefits on regional territory. A policy maker might consider a re-allocation of resources among sectors, or some stimulus for local enterprise in this particular sector. Subsequent analysis (Table 3) will actually show that universities in 3 of these regions (Friuli Venezia-Giulia, Marche and Trentino Alto Adige) collaborate with extra-regional enterprises. Column 4 illustrates the degree of correspondence between supply and demand, expressed as the difference between the number of local university scientists and the number of collaborations activated by local enterprises (the "surplus"). Lombardy is the only region observed to have a shortage of supply, and it is a strong shortage (-32), while strong surpluses are seen for Lazio (+40), Emilia Romagna (+31), Tuscany (+26) and Campania (+22). Column 5 presents results for the ratio method of analyzing correspondence in each region, expressed as the national demand for collaboration by local enterprises per university scientist in the region. Situations of high surplus will likely signal the occurrence of a high market share for local universities. In contrast, regions characterized by marked shortages of supply should find that the market share of local universities is quite low. To better enable comparisons between regions, the degree of supply-demand correspondence has also been expressed relative to the mean value of the distribution of regional scores. For example, the value 6.61 for Lombardy (seen in parentheses, column 4), indicates that the industrial demand per university scientist is 6.6 times more than the mean value of the national distribution (0.25). Two other regions result as having values above the mean: Sicily (2.81) and Abruzzo (3.26).

---

[8] The assumption can easily be modified to adapt the analysis to the characteristics of different SDSs, or in light of the personnel resources that universities might assign to respond to industrial demand for collaboration.

[9] This interpretation is not intended as a superficial suggestion that universities should resize their research capacity in the SDS examined. Capacity must also be planned in relation to the other two primary roles of the university: higher education and research.

To appreciate the potential of the tool for assessing the impact of policies, for the incumbent regional policy maker at time $t_o$, the degree of correspondence at the same time $t_o$ could be viewed as a reference point. At a subsequent time $t_1$, the measure of variation in degree of correspondence could then reveal the effectiveness or ineffectiveness of policies and actions undertaken to align demand and supply of new knowledge.

**5.2 Regional appropriability of academic knowledge spillovers**

The regional policy maker is particularly interested in maximizing economic benefits from the public supply new knowledge within her/his region. Confronted with a regional demand for new knowledge she/he is interested in knowing if, and in what measure, local supply will satisfy the demand; also in knowing what knowledge from local universities may be migrating towards "competing" regions, rather than remaining to satisfy local demand. The greater the level of internal demand that is satisfied by local supply, the lesser will be the transaction costs due to geographic distance, for local enterprises. In the situation of unsatisfied internal demand, the less will be the extra-regional collaboration by local universities, along with a lesser probability for boomerang effects on regional competitivity in the sector. Table 3 presents regional flows in research collaboration, again as seen in the Electronics SDS.

[Table 3]

Regions characterized by universities with a strong habit of extra-regional collaboration are, in order, (from column 4): Veneto (1.75), Lombardy (0.830) and Umbria (0.80). Regions with a strong shortage or low surplus of supply should show a high value of intra-regional supply per university scientist: Lombardy (0.70), Abruzzo (0.60) and Sicily (0.38) show the highest values (column 6), confirming the expectation.

Columns 7 and 8 show, respectively, the regional market share of the local universities (number of university-enterprise collaborations divided by total number of collaborations undertaken by enterprises of the region) and the regional market share per single scientist. It can immediately be seen that Piedmont, Campania and Liguria, while presenting a significant supply, do not undertake any intra-regional collaborations. Further, in Piedmont and Liguria the universities also do not undertake any extra-regional collaborations, while in Campania the extra-regional spillovers are significant, in spite of there also being an internal demand. In 4 regions (Abruzzo, Puglia, Umbria and Veneto) the internal demand is entirely satisfied by the local university supply. For the regional market share per university scientist, Abruzzo and Umbria stand apart from all other regions, managing to satisfy demand in full in spite of a very low surplus with respect to other regions (individual scientists with 20% of market share). Meanwhile, Lombardy presents a very low share per scientist (0.89%), confirming the expectations given the elevated demand with respect to the supply.

Analysis of local university collaboration with extra-regional enterprises shows the extent of external spillovers, something that would be particularly undesirable in cases where local enterprises are competing along the same technological trajectory with extra-regional enterprises in the same field. From the opposite perspective, such analysis reveals the potential of a region to attract extra-regional investments in a given technological dimension.

The last indicator proposed (column 9) represents the ratio of the collaborations realized by universities within their region to the total of their collaborations in the nation as a whole, which indicates how much of their new knowledge produced provides benefits within the region compared to how much goes to the nation as a whole, and thus to the "competitor" regions. Abruzzo, Sicily, and Lombardy are the three regions characterized by the highest appropriation of knowledge from local universities by their local enterprises compared to the appropriation by extra-regional enterprises (intra-regional supply/national supply). The ratio for these three regions is, respectively, 100%, 88.9%, 84.62%. At the lowest level of the classification are Veneto (4.76%) and Campania (0.00%).

**5.3 A diagnostic tool**

The preceding analyses give an example of measurement of degree of correspondence and the flows of university-industry collaboration for a single SDS. Analysis of historical data would permit measurement of the efficacy of policies for regional development, particularly policies intending to increase the degree of correspondence between public and private research and the interaction between these two spheres in a selected SDS. The regional policy maker could take advantage of the proposed methodology to compare the positioning of her/his own region in comparison to others, along various dimensions. As an example, Figure 1 shows, for the Electronics SDS, the position of each region along two dimensions: i) degree of correspondence between demand and supply; ii) capacity to satisfy internal demand (regional market share). The figure shows only regions that present a demand for research collaboration in Electronics. The regions best positioned in terms of correspondence are those found around the vertical line that divides the quadrants. Abruzzo, Umbria and Sicily are the regions closer to this line. Those with the best capacity of satisfying internal demand with local university resources are positioned in quadrants I and II. Any "virtuous" regions (none occur in this example) would be located in quadrant I, succeeding in satisfying a high level of internal demand even though in a situation of shortage. On the opposite side, quadrant III contains those regions that, even with ample potential (surplus of supply), do not succeed in delivering knowledge produced to their local enterprises. In the example here, all the regions but one are situated in quadrants II and III. The only exception, Lombardy (region ID: 6), is situated in quadrant IV. Though confronted with a high shortage in terms of degree of correspondence (-32), universities in Lombardy succeed in meeting 40% of the knowledge demand from local private enterprises. The regions that succeed in saturating internal demand are Abruzzo, Umbria, Veneto and Puglia. Abruzzo and Umbria can be considered the most virtuous regions, considering their minimal surplus of supply. Campania, Piedmont and Liguria offer the contrast of regions that, even with a significant surplus, do not activate any intra-regional collaborations. The supply from Lazio, which represents the highest surplus of all (40), captures only 23% of local demand, the lowest score after the three above-noted regions with no intra-regional collaborations.

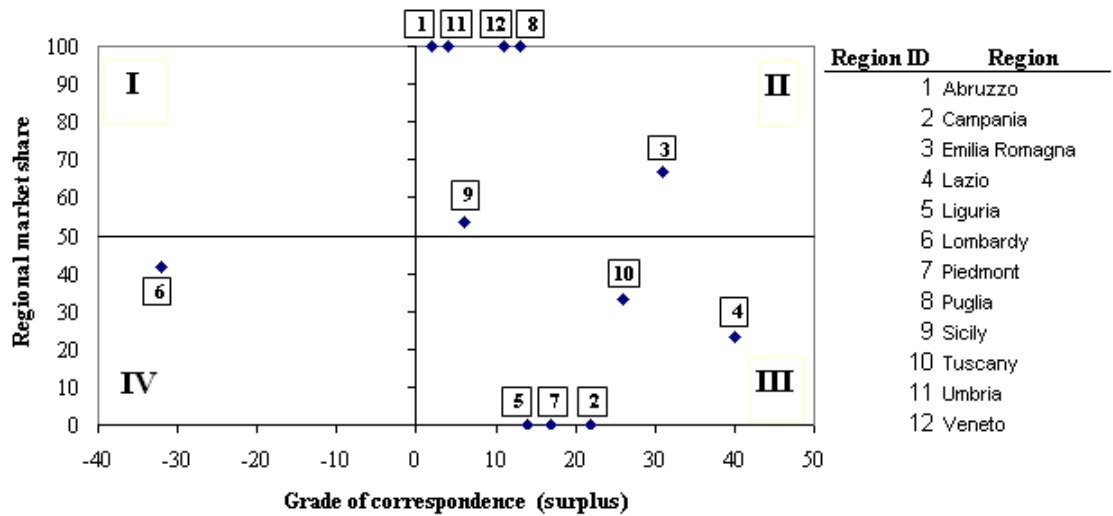

*Figure 1: Positioning of Italian regions in terms of degree of correspondence and regional market share for the Electronics SDS*

Another possible application of the tool is to compare all the SDS for a single region in a single dimension. As an example, Table 4 shows the analysis of degree of correspondence for Lombardy in all the SDS with incumbent university personnel. Lombardy is notable in Italy for its high level of academic supply and, above all, for its high concentration of private research. (For example, referring to Table 3 and the Electronics SDS, private enterprises in Lombardy contribute, in terms of SDS-enterprise collaborations in Italy, roughly 59% of the entire national demand for knowledge).

[Table 4]

The range of values for degree of correspondence (national demand per university scientist) in Lombardy for all the SDS considered is quite broad: from a zero value in a full 64 SDS with at least one university researcher, to a maximum value of 4.778 for the SDS "Industrial chemistry".

In the light of this and possible other information, the regional policy maker can identify relative imbalances between SDS in their degree of demand-supply correspondence. Following the "smart specialization" policy recently pursued by the EU (McCann and Ortega-Argilés, 2011), in keeping with the strategies for regional industrial development, she/he can then focus on interventions to shift resources from SDS with a surplus supply to those with a shortage, or to stimulate birth of spin-off enterprises and increase extra-regional greenfield investment in the SDS with a surplus of supply.

In cases such as Italy, where the university system is almost entirely public and, at a financial level, highly dependent on central government, the regional policy maker's possibilities for maneuver are quite limited[10]. For the regions of Italy, the degree of correspondence in demand-supply should be interpreted as a structural factor and not as an indicator of virtuosity.

## 6. Regional aggregate analysis

To obtain an ordering of the relative overall level of virtuosity of each region, analysis proceeded to the weighed aggregation for data at the sectorial level[11]. Each sectorial level indicator was weighed according to relative importance of the SDS versus the total of SDS in which collaborations occur. Table 5 illustrates the results of this aggregation process.

[Table 5]

At this level of aggregated data, the degree of correspondence between demand and supply was measured through national demand per university scientist (column 2). However, any interpretation of this aggregate level data must be very cautious. In fact, a region with a strong surplus in one sector and an equivalent weight of shortage in another sector would indicate a zero net balance. While noting this caution, the regions that present the highest values, noticeably differentiated from the others, are seen as Trentino Alto Adige (0.831), Lombardy (0.774), Friuli VeneziaGiulia (0.516) and Piedmont (0.416). Those that register an excess of supply with respect to demand are Basilicata (which shows a 0 value due to the lack of demand for collaboration on the part of local private enterprises), Puglia (0.074) and Umbria (0.115). We would expect that university researchers in these last two regions would succeed in meeting the needs of local enterprises (regional market share per university scientist, column 5) to a greater extent than their colleagues in other regions. In effect, Puglia is situated in seventh place (4.61%) and Umbria in fifth (9.87%). The top three placements are Marche (11.50%), Campania (11.31%) and Abruzzo (11.18%), which are all regions of south-central Italy. These regions thus result as those in which universities best succeed at meeting the needs of local private enterprises.

At the level of supply per university scientist, considering collaborations by university researchers of a particular region with private enterprises in all Italy (i.e. national supply per university scientist, column 3), the top-ranking three regions are Trentino Alto Adige (1.223), Umbria (0.999) and Calabria (0.812). Considering collaborations with only local industry (intra-regional supply per university scientist, column 4), the highest values are observed for Lombardy (134.52), Abruzzo (64.487)

---

[10] However, over time, there have been significant delegations of central authority to the regions (Title V of the constitution), and these are tending to increase, resulting in a current gradual development of regional federalism. For example, the regions have recently obtained the power to enact incentive measures for research in specific sectors, with appropriately targeted financing.

[11] See Abramo et al. (2008) for further information on potential distortions in aggregate bibliometric analyses that do not consider sectorial specificity.

and Friuli VeneziaGiulia (61.599).

For the last indicator proposed, which is intra-regional supply/national supply (column 6), the top-ranked regions result as Lombardy (0.584), Sicily (0.537), Piedmont (0.425), Lazio (0.388) and Emilia Romagna (0.316). In these regions, the quantity of public knowledge produced *in loco* that remains in the region, compared to that which departs, is highest. Apart from the island region of Sicily, where difficult connections with the rest of Italy undoubtedly play a relevant penalizing role, these are all regions with a high industrial concentration, thus characterized by a high level of demand.

## 7. Conclusions

The present work proposes a complementary support instrument for regional policy makers, to assist them in identifying and quantifying the sectorial demand and supply of knowledge in the market for public-private research collaborations. To measure research collaborations the study has adopted a bibliometric approach, based on co-authorship of international journal articles. The approach takes into consideration the production of scientific publications, and is a valid complement to investigations based on patents, which in the Italian case result as extremely limited and thus poorly representative. The instrument proposed permits numerous possible applications by the regional policy maker. Above all, it provides a measure of the degree of correspondence between demand and supply of knowledge, meaning the extent of alignment in the directions of public and private research in a particular region. To this end, each Italian region has been characterized in terms of demand for new knowledge (by private enterprises) and supply of the same (from universities), at three levels: intra-regional (between local partners), extra-regional (between partners from different regions) and national (between both local and extra-regional partners). Such cognitive description of the scientific-technological structure of a region (the "as is" situation) is an indispensable requisite for formulation of policies directed at improving structural correspondence (the "to be" situation). Further, through time-series analyses of evolution in collaboration, it is possible to evaluate the effectiveness of actions undertaken to increase the intensity of public-private collaboration in a region, focusing on sectors identified as strategic for the regional development plan. The instrument also permits measurement of the relative capacity of the region to maximize, through public-private collaboration, the local economic benefits arising from new knowledge produced by its own universities, both in each technical-scientific sector and for the aggregate level. In fact, a regional policy maker aims to maximize retention in her/his region of the benefits from the local university supply of new knowledge, and thus, facing a regional demand for new knowledge, she/he is interested in knowing if and in what measure such demand is satisfied by local supply and then how much of any knowledge produced by local universities actually migrates towards "competitor" regions, rather than satisfying internal demand.

For purposes of illustration, this work furnishes several examples of application of the tool to the Italian case: a description and comparison of intensity of public-private collaborations in a specific SDS (Electronics) for the various regions; the analysis, for one region (Lombardy), of the collaborations in the various SDS; comparison between the regions, again in the Electronics SDS, in terms of degree of correspondence and appropriability for public knowledge spillovers with respect to local industry.

The results obtained and analyzed through this kind of application, in conjunction with other diagnostic analyses using complementary tools, can provide useful policy indications at diverse levels of intervention. After taking into account for the specific characteristics of each regional context, such results can stimulate a more effective and efficient use of resources present in a region, both at the research institutions and industry levels: for example in re-balancing, where appropriate considering all other interplaying constraints, the sectorial supply also in relation to demand, or fostering the launch of enterprises in sectors with strong supply. Needless to say that the principle universities' institutional mission is education and the availability of competences in specific fields of knowledge may simply respond to educational purposes, regardless of a regional demand by industry. The results of analysis can also motivate targeted sectorial interventions, intended to favour increasing use of university research results on the part of the industrial sector, with evident benefits for the innovation processes and economic development of the region. A potential caveat in the application of the tool descends from the association of the SDS sector of collaboration to the researcher. However, researchers can move across universities, especially at the beginning of their career. Associating a structural policy, such as the sectorial policy, to a mobile factor can raise some problem. However, research staff shifts entail most of the times replacements by other researchers in the same field, to fulfill educational needs. Furthermore, in some national contexts (the Italian one, for example) mobility across universities in the period considered is not so relevant and may concern one or a few units within a large group (SDS).

Future developments of this work could include the application of bibliometric network analysis through mapping and clustering techniques which have flourished in recent years (Waltman et al., 2010; Leydesdorff and Rafols, 2009; Noyons and Calero-Medina, 2009; Boyak et al., 2005). Parallel to this, sophisticated software for elaboration and visualization of such networks could be applied (Van Eck and Waltman, 2010). In terms of scope and content, the application of the described approach could be extended to other nations, and also include data on patent production, or the analysis of the Italian case could be extended to cover a longer time interval (preferably adding the analysis of the supply of knowledge from public research institutes other than universities). Another interesting development, on which the authors are already working, concerns the identification of the principal determinants (geographic distance, presence of star scientists, etc.) and their relative weight, in the phenomenon of collaborations between universities and industry.

**Acknowledgments**

The authors express their sincere thanks to Flavia Di Costa, for her invaluable contribution to the data analysis. Any possible inaccuracies or other errors remain as the complete responsibility of the authors.

|  | Supply of collaboration[*] | | | Demand for collaboration[**] | | | Net difference[***] | Regional market share (%)[****] |
|---|---|---|---|---|---|---|---|---|
| Region | Intra-regional | Extra-regional | National (total) | Intra-regional | Extra-regional | National (total) | | |
| Abruzzo | 13 | 44 | 57 | 13 | 10 | 23 | 34 | 57 |
| Basilicata | 0 | 6 | 6 | 0 | 0 | 0 | 6 | NA |
| Calabria | 0 | 13 | 13 | 0 | 2 | 2 | 11 | 0 |
| Campania | 13 | 90 | 103 | 13 | 8 | 21 | 82 | 62 |
| Emilia Romagna | 93 | 205 | 298 | 93 | 103 | 196 | 102 | 47 |
| Friuli Venezia-Giulia | 15 | 45 | 60 | 15 | 6 | 21 | 39 | 71 |
| Lazio | 63 | 97 | 160 | 63 | 226 | 289 | -129 | 22 |
| Liguria | 7 | 52 | 59 | 7 | 16 | 23 | 36 | 30 |
| Lombardy | 233 | 170 | 403 | 233 | 536 | 769 | -366 | 30 |
| Marche | 6 | 31 | 37 | 6 | 2 | 8 | 29 | 75 |
| Molise | 0 | 0 | 0 | 0 | 2 | 2 | -2 | 0 |
| Piedmont | 57 | 77 | 134 | 57 | 90 | 147 | -13 | 39 |
| Puglia | 3 | 45 | 48 | 3 | 1 | 4 | 44 | 75 |
| Sardinia | 3 | 22 | 25 | 3 | 6 | 9 | 16 | 33 |
| Sicilia | 62 | 56 | 118 | 62 | 23 | 85 | 33 | 73 |
| Toscana | 67 | 146 | 213 | 67 | 148 | 215 | -2 | 31 |
| Trentino Alto Adige | 2 | 18 | 20 | 2 | 8 | 10 | 10 | 20 |
| Umbria | 3 | 56 | 59 | 3 | 1 | 4 | 55 | 75 |
| Veneto | 50 | 120 | 170 | 50 | 105 | 155 | 15 | 32 |

*Table 1: Regional distribution of research collaborations, 2001-2003: distribution of demand and supply*

[*] "Supply of collaboration" (intra-regional/extra-regional/national) for any region is the number of university-enterprise collaborations of the universities of that local region with private enterprises in intra-regional, extra-regional and national locations

[**] "Demand for collaboration" (intra-regional/extra-regional/national) for a region is the number of university-enterprise collaborations of the private enterprises of that local region with universities in intra-regional, extra-regional and national locations

[***] Net difference in a region is calculated as national supply of collaboration (from the region) less national demand for collaboration in the same region.

[****] "Regional market share" is defined as the ratio of the number of collaborations of local universities with local enterprises, respective to the national demand for collaboration by local enterprises (i.e. the ratio of supply of intra-regional collaboration to national demand for collaboration).

| Region | Number of university scientists (% of total) | National demand* (% of total) | Degree of correspondence | |
|---|---|---|---|---|
| | | | Surplus** | National demand per university scientist (difference from national mean) |
| Abruzzo | 5 (1.56) | 3 (2.24) | 2 | 0.60 (2.36) |
| Basilicata | 0 (0) | 0 (0.00) | 0 | NA |
| Calabria | 6 (1.88) | 0 (0.00) | 6 | 0.00 (0.00) |
| Campania | 24 (7.50) | 2 (1.49) | 22 | 0.08 (0.33) |
| Emilia Romagna | 37 (11.56) | 6 (4.48) | 31 | 0.16 (0.64) |
| Friuli Venezia-Giulia | 12 (3.75) | 0 (0.00) | 12 | 0.00 (0.00) |
| Lazio | 53 (16.56) | 13 (9.70) | 40 | 0.25 (0.96) |
| Liguria | 15 (4.69) | 1 (0.75) | 14 | 0.07 (0.26) |
| Lombardy | 47 (14.69) | 79 (58.96) | -32 | 1.68 (6.61) |
| Marche | 3 (0.94) | 0 (0.00) | 3 | 0.00 (0.00) |
| Molise | 0 (0) | 0 (0.00) | 0 | NA |
| Piedmont | 26 (8.13) | 6 (4.48) | 17 | 0.23 (0.91) |
| Puglia | 14 (4.38) | 1 (0.75) | 13 | 0.07 (0.28) |
| Sardinia | 6 (1.88) | 0 (0.00) | 6 | 0.00 (0.00) |
| Sicily | 21 (6.56) | 15 (11.19) | 6 | 0.71 (2.81) |
| Tuscany | 32 (10.00) | 6 (4.48) | 26 | 0.19 (0.74) |
| Trentino Alto Adige | 2 (0.63) | 0 (0.00) | 2 | 0.00 (0.00) |
| Umbria | 5 (1.56) | 1 (0.75) | 4 | 0.20 (0.79) |
| Veneto | 12 (3.75) | 1 (0.75) | 11 | 0.08 (0.33) |

*Table 2: Degree of correspondence between industry demand and potential university supply of research collaboration in electronics*

\* "National demand" of a region indicates the number of collaborations undertaken by its local enterprises with universities located anywhere in Italy.

\*\* "Surplus" of a region is defined as the difference between the number of collaborating scientists from local universities and the number of collaborations undertaken by the private enterprises of the same region.

| Region | National demand (% of total national demand) | National supply* (% of total) | National supply per university scientist (difference from national mean) | Intra-regional supply** (% of total) | Intra-regional supply per university scientist (difference from national mean) | Regional market share (%)*** | Regional market share (%) per university scientist | Intra-regional supply / National supply (%) |
|---|---|---|---|---|---|---|---|---|
| Abruzzo | 3 (2.24) | 3 (2.24) | 0.60 (1.39) | 3 (2.24) | 0.60 (4.50) | 100.00 | 20.00 | 100.00 |
| Basilicata | 0 (0.00) | 0 (0.00) | NA | NA | NA | NA | NA | NA |
| Calabria | 0 (0.00) | 0 (0.00) | 0.00 (0.00) | NA | NA | NA | NA | NA |
| Campania | 2 (1.49) | 4 (2.99) | 0.17 (0.39) | 0 (0.00) | 0.00 (0.00) | 0.00 | 0.00 | 0.00 |
| Emilia Romagna | 6 (4.48) | 22 (16.42) | 0.59 (1.38) | 4 (2.99) | 0.11 (0.81) | 66.67 | 1.80 | 18.18 |
| Friuli Venezia-Giulia | 0 (0.00) | 3 (2.24) | 0.25 (0.58) | NA | NA | NA | NA | NA |
| Lazio | 13 (9.70) | 13 (9.70) | 0.25 (0.57) | 3 (2.24) | 0.06 (0.42) | 23.08 | 0.44 | 23.08 |
| Liguria | 1 (0.75) | 0 (0.00) | 0.00 (0.00) | 0 (0.00) | 0.00 (0.00) | 0.00 | 0.00 | NA |
| Lombardy | 79 (58.96) | 39 (29.10) | 0.83 (1.92) | 33 (24.63) | 0.70 (5.27) | 41.77 | 0.89 | 84.62 |
| Marche | 0 (0.00) | 1 (0.75) | 0.33 (0.77) | N.A. | NA | NA | NA | NA |
| Molise | 0 (0.00) | 0 (0.00) | NA | N.A. | NA | NA | NA | NA |
| Piedmont | 6 (4.48) | 0 (0.00) | 0.00 (0.00) | 0 (0.00) | 0.00 (0.00) | 0.00 | 0.00 | NA |
| Puglia | 1 (0.75) | 10 (7.46) | 0.71 (1.65) | 1 (0.75) | 0.07 (0.54) | 100.00 | 7.14 | 10.00 |
| Sardinia | 0 (0.00) | 0 (0.00) | 0.00 (0.00) | N.A. | NA | NA | NA | NA |
| Sicily | 15 (11.19) | 9 (6.72) | 0.43 (0.99) | 8 (5.97) | 0.38 (2.86) | 53.33 | 2.54 | 88.89 |
| Tuscany | 6 (4.48) | 4 (2.99) | 0.13 (0.29) | 2 (1.49) | 0.06 (0.47) | 33.33 | 1.04 | 50.00 |
| Trentino Alto Adige | 0 (0.00) | 1 (0.75) | 0.50 (1.16) | NA | NA | NA | NA | NA |
| Umbria | 1 (0.75) | 4 (2.99) | 0.80 (1.85) | 1 (0.75) | 0.20 (1.50) | 100.00 | 20.00 | 25.00 |
| Veneto | 1 (0.75) | 21 (15.67) | 1.75 (4.05) | 1 (0.75) | 0.08 (0.63) | 100.00 | 8.33 | 4.76 |

*Table 3: Regional flows in research collaboration in electronics*
\* "National supply" for a region indicates the number of collaborations by local universities with private enterprises located anywhere in Italy
\*\* "Intra-regional supply" of a region indicates the number off collaborations by local universities with local private enterprises.
\*\*\* "Regional market share" of local universities is defined as the ratio between number of collaborations by local universities with local enterprises (intra-regional supply) and the number of collaborations undertaken by local private enterprises with universities located anywhere in Italy (national demand).

| | | |
|---|---|---|
| National demand per university scientist | Observations* | 173 |
| | Mean | 0.227 |
| | Standard error | 0.039 |
| | Median | 0.083 |
| | Minimum | 0 |
| | Maximum | 4.778 |
| | Number of SDS with no national demand | 64 |

*Table 4: Statistics for degree of coincidence in the region of Lombardy*
\* The analysis does not include the 10 SDS without incumbent university researchers

| Region | National demand per university scientist (Ranking) | National supply per university scientist (Ranking) | Intra-regional supply per university scientist (Ranking) | Regional market share (%) per university scientist (Ranking) | Intra-regional supply / National supply (Ranking) |
|---|---|---|---|---|---|
| Abruzzo | 0.351 (6) | 0.553 (4) | 0.451 (2) | 11.180 (3) | 0.258 (8) |
| Basilicata | 0.000 (19) | 0.349 (9) | 0.000 (17) | N.A (N.A.) | N.A (N.A.) |
| Calabria | 0.375 (5) | 0.812 (3) | 0.000 (17) | 0.000 (17) | 0.000 (17) |
| Campania | 0.160 (13) | 0.175 (16) | 0.210 (6) | 11.308 (2) | 0.138 (12) |
| Emilia Romagna | 0.241 (11) | 0.341 (10) | 0.162 (9) | 2.095 (12) | 0.316 (5) |
| Friuli Venezia-Giulia | 0.516 (3) | 0.493 (6) | 0.562 (1) | 10.269 (4) | 0.257 (9) |
| Lazio | 0.235 (12) | 0.137 (18) | 0.104 (15) | 0.822 (16) | 0.388 (4) |
| Liguria | 0.135 (15) | 0.514 (5) | 0.128 (12) | 3.930 (9) | 0.123 (13) |
| Lombardy | 0.774 (2) | 0.258 (13) | 0.222 (5) | 1.077 (15) | 0.584 (1) |
| Marche | 0.155 (14) | 0.215 (15) | 0.158 (10) | 11.505 (1) | 0.160 (11) |
| Molise | 0.250 (10) | 0.000 (19) | 0.000 (17) | 0.000 (17) | 0.000 (17) |
| Piedmont | 0.416 (4) | 0.393 (8) | 0.199 (7) | 2.001 (14) | 0.425 (3) |
| Puglia | 0.074 (18) | 0.280 (11) | 0.077 (16) | 4.614 (7) | 0.057 (16) |
| Sardinia | 0.127 (16) | 0.170 (17) | 0.110 (14) | 5.060 (6) | 0.207 (10) |
| Sicily | 0.325 (8) | 0.216 (14) | 0.232 (4) | 3.020 (10) | 0.537 (2) |
| Tuscany | 0.287 (9) | 0.268 (12) | 0.132 (11) | 2.061 (13) | 0.310 (6) |
| Trentino Alto Adige | 0.831 (1) | 1.223 (1) | 0.250 (3) | 4.545 (8) | 0.077 (14) |
| Umbria | 0.115 (17) | 0.999 (2) | 0.123 (13) | 9.870 (5) | 0.058 (15) |
| Veneto | 0.339 (7) | 0.435 (7) | 0.000 (17) | 2.101 (11) | 0.302 (7) |

*Table 5: Principal aggregated indexes (for all SDS considered) of SDS-enterprise collaborations*